\documentclass[preprint,showpacs]{revtex4}
\usepackage{graphicx}
\usepackage{amsmath,amssymb,amsfonts}
\usepackage{floatflt}
\usepackage{wrapfig}
\usepackage{indentfirst}

\begin{document}

\title{The quantization of persistent current qubit. The role of inductance}
\author{Ya. S. Greenberg}
\affiliation{Novosibirsk State Technical University, 20 K. Marx
Ave., 630092 Novosibirsk, Russia}

\date{\today}

\pacs{03.67.Lx
, 85.25.Cp
, 85.25.Dq
, 85.35.Ds}
\begin{abstract}

The Hamiltonian of persistent current qubit is found within well
known quantum mechanical procedure. It allows a selfconsistent
derivation of the current operator in a two state basis. It is
shown that the current operator is not diagonal in a flux basis. A
non diagonal element comes from the finite inductance of the
qubit. The results obtained in the paper are important for the
circuits where two or more flux qubits are coupled inductively.

\end{abstract}

\maketitle
\section{Introduction}
Josephson-junction qubits are known to be candidates for scalable
solid-state quantum computing circuits \cite{Makhlin}. Here we
consider a supercobducting flux qubit which has been first
proposed in \cite{Mooij} and analyzed in \cite{Orlando} and
\cite{Mooij1}. The qubit consists of three Josephson junctions in
a loop with very small inductance~$L$, typically in the pH range.
This insures effective decoupling from the environment. However,
in the practical implementation of flux qubit circuitry it is
important to have the loop inductance as much as possible
consistent with a proper operation of a qubit. A relative large
loop inductance facilitates a qubit control biasing schemes and
the formation, control and readout of two-qubit quantum gates.
These considerations stimulated some investigations of the role
the loop inductance plays in the operation of a flux qubit
\cite{Crankshaw}, \cite{Brink}, \cite{You}. The main goal of these
works was the calculation of the corrections to the energy levels
due to finite inductance of the loop. In the early work
\cite{Crankshaw} these corrections have been obtained by
perturbation expansion of the energy over small parameter
$\beta=L/L_J$, where $L_J$ is the Josephson junction inductance.
The extension to large $\beta$'s (up to $\beta\approx 10$) had
been considered in \cite{You}. However, it is important to realize
that for finite loop inductance the interaction between two state
qubit with its own LC circuit cannot in general be neglected. If
$\beta$ is not small, as in \cite{You}, this interaction can have
substantial influence on the energy levels. Unfortunately, this
interaction in \cite{You} has been completely neglected.

In principle, the account for a finite loop inductance (even if it
is small) requires the correct construction of quantum mechanical
Hamiltonian of a qubit, which contains all relevant interactions.
This has been done in \cite{Brink}, where the effective
Hamiltonian has been obtained by a rigorous expansion procedure in
powers of $\beta$. As was shown in \cite{Brink}, one of the effect
of the interaction of  a flux qubit with its own LC oscillator is
the renormalization of the Josephson critical current. The
inclusion of circuit inductances in a systematic derivation of the
Hamiltonian of superconducting circuits has been done in
\cite{Bur1}. It allows the correct calculations of the effects of
the finite inductance both for flux \cite{Bur2} and charge
\cite{Bur3} qubits.

In this paper we investigate another physical effect which comes
from finite loop inductance. Namely, we show that the finite loop
inductance results in the additional term of the current operator
in the flux basis:
\begin{equation}\label{Cop}
    \widehat{I}=A\tau_Z+B\tau_X,
\end{equation}
where $\tau_Z$ and $\tau_X$ are Pauli matrices in the flux basis.
The quantities $A$ and $B$ in (\ref{Cop}) are calculated in the
paper the $B$ being conditioned by the finite loop inductance: for
$L=0$ the second term in (\ref{Cop}) is absent. Though for the
usual qubit design with small loop inductance this second term is
relatively small, nevertheless, it might give noticeable effects
for large $\beta$'s for the arrangements when two flux qubit are
coupled either via a common inductance \cite{You} or inductively
coupled via a term $M\widehat{I}_1\widehat{I}_2$ in the
Hamiltonian, where $M$ is a mutual inductance between qubit's
loops, $\widehat{I}_1$, $\widehat{I}_2$ are the current operators
of the respective qubits.

The paper is organized as follows. In Section II we start with the
exact Lagrangian of a flux qubit with finite loop inductance. In
section III with the aid of well known procedure we derive
rigorously the quantum qubit Hamiltonian. The current operator is
studied in Section IV, where we show that in general it is not
diagonal in the flux basis. The matrix elements for the current
operator are calculated in Section VI, where in order to obtain
analytical results we consider a flux qubit with a small loop
inductance.

\section{Lagrangian for the flux qubit}

\begin{figure}
  \includegraphics[width=8 cm, angle=-90]{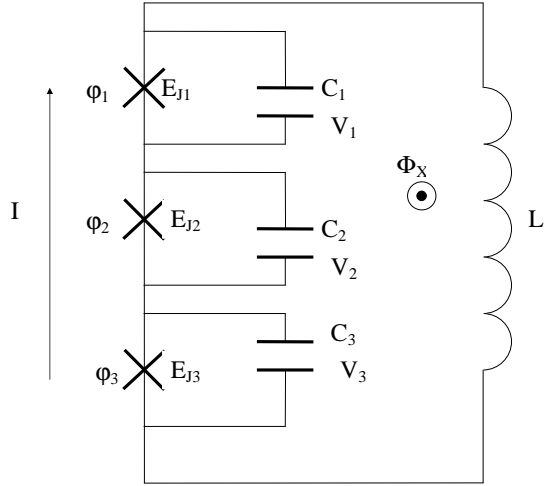}
  \caption{A flux qubit, where an external magnetic flux $\Phi_X$ pierces the
  superconducting loop that contains three Josephson junctions and inductance $L$.
  Two Josephson junctions are considered to be identical, $E_{J1}=E_{J2}=E_J$,
  $C_1=C_2=C$, and $E_{J3}=\alpha E_J$, $C_3=\alpha C$.}\label{Fig1}
\end{figure}

We consider here a well known design of the flux qubit with three
Josephson junctions \cite{Mooij}, \cite{Orlando}, \cite{Mooij1},
which is shown on Fig.\ref{Fig1}.

Two junctions have equal critical current $I_\mathrm{c}$ and
(effective) capacitance~$C$, while those of the third junction are
slightly smaller: $\alpha I_\mathrm{c}$ and $\alpha C$, with
$0.5<\alpha<1$. If the Josephson energy
$E_\mathrm{J}=I_\mathrm{c}\Phi_0/2\pi$ is much larger than the
Coulomb energy $E_C=e^2\!/2C$, the Josephson phase is well
defined. Near $\Phi_\mathrm{x}=\Phi_0/2$, this system has two
low-lying quantum states \cite{Orlando, Mooij1}. The Lagrangian of
this qubit is the difference between the charge energy in the
junction capacitors and the sum of Josephson and magnetic energy:

\begin{equation}\label{Lag1}
    L = \sum\limits_{i = 1}^3 {\frac{{C_i V_i^2 }}{2} + } \sum\limits_{i = 1}^3 {E_J^{(i)}
     \cos \varphi _i  - \frac{{\Phi ^2 }}{{2L}}}
\end{equation}

where $V_i$ is the voltage across the junction capacitance $C_i$,
which is related to the phase $\phi_i$ by the Josephson relation
$V_i=(\Phi_0/2\pi) \dot{\phi_i}$; $\Phi$ is the flux trapped in
the loop:

\begin{equation}\label{Flux}
    \Phi  = \frac{{\Phi _0 }}{{2\pi }}\sum\limits_{i = 1}^3 {\varphi _i  - \Phi _X }
\end{equation}
Next we make the following definitions:
$\phi=\phi_1+\phi_2+\phi_3$, $\phi_1+\phi_2=2\theta$,
$\phi_1-\phi_2=2\chi$. In terms of these new phases Lagrangian
(\ref{Lag1}) takes the form:
\begin{equation}\label{Lag2}
 L = \frac{{\hbar ^2 }}{{16E_C }}\left( {\dot \varphi _1^2  + \dot \varphi _2^2 }
     \right) + \alpha \frac{{\hbar ^2 }}{{16E_C }}\dot \varphi _3^2  +
     2E_J \cos \theta \cos \chi  + \alpha E_J \cos \left( {\varphi  - 2\theta }
      \right) - \frac{{E_J }}{{2\beta }}\left( {\varphi  - \varphi _X } \right)^2
\end{equation}
where $\beta=2\pi LI_C/\Phi_0$.
\section{Construction of Hamiltonian}
 Conjugate variables are defined in a standard way:
\begin{equation}\label{nphi}
n_\varphi   = \frac{1}{\hbar }\frac{{\partial L}}{{\partial \dot
\varphi }} = \alpha \frac{\hbar }{{8E_C }}\left( {\dot \varphi  -
2\dot \theta } \right)
\end{equation}

\begin{equation}\label{ntet}
n_\theta   = \frac{1}{\hbar }\frac{{\partial L}}{{\partial \dot
\theta }} = \frac{\hbar }{{4E_C }}\dot \theta  - \alpha
\frac{\hbar }{{4E_C }} \left( {\dot \varphi  - 2\dot \theta }
\right)
\end{equation}

\begin{equation}\label{nhi}
n_\chi   = \frac{1}{\hbar }\frac{{\partial L}}{{\partial \dot \chi
}} = \frac{\hbar }{{4E_C }}\dot \chi
\end{equation}
From these equations we express phases in terms of conjugate
variables:
\begin{equation}\label{Motion1}
\dot \varphi  = \frac{{8E_C }}{\hbar }\frac{{2\alpha  + 1}}{\alpha
}n_\varphi   + \frac{{8E_C }}{\hbar }n_\theta
\end{equation}

\begin{equation}\label{Motion2}
\dot \theta  = \frac{{4E_C }}{\hbar }n_\theta   + \frac{{8E_C
}}{\hbar }n_\varphi
\end{equation}

\begin{equation}\label{Motion3}
\dot \chi  = \frac{{4E_C }}{\hbar }n_\chi
\end{equation}
In terms of conjugate variables Lagrangian (\ref{Lag2}) takes the
form:
\begin{eqnarray}\label{Lag3}
L = 2E_C n_\theta ^2  + 2E_C n_\chi ^2  + 4E_C \frac{{1 + 2\alpha
}}{\alpha }n_\varphi ^2  + 8E_C n_\theta  n_\varphi\\\nonumber
+2E_J \cos \theta \cos \chi  + \alpha E_J \cos \left( {\varphi  -
2\theta } \right) - \frac{{E_J }}{{2\beta }}\left( {\varphi  -
\varphi _X } \right)^2
\end{eqnarray}
The Hamiltonian is constructed according to the well known rule:
\begin{equation}\label{Ham1}
H = \hbar n_\varphi  \dot \varphi  + \hbar n_\theta  \dot \theta +
\hbar n_\chi  \dot \chi  - L
\end{equation}
Finally we obtain:
\begin{equation}\label{Ham2}
H = 4E_C \frac{{2\alpha  + 1}}{\alpha }n_\varphi ^2  + 2E_C n_\chi ^2  + 2E_C n_\theta ^2
 + 8E_C n_\theta  n_\varphi+U(\chi, \theta, \varphi)
\end{equation}
where
\begin{equation}\label{PotEn}
    U(\chi, \theta, \varphi)= - 2E_J \cos \theta \cos \chi  - \alpha E_J \cos
 \left( {\varphi  - 2\theta } \right) + \frac{{E_J }}{{2\beta }}\left( {\varphi  -
 \varphi _X } \right)^2
\end{equation}

Hence, equations of motion for the phases (\ref{Motion1}),
(\ref{Motion2}), (\ref{Motion3}) are simply $\dot \varphi  =
\frac{1}{\hbar }\frac{{\partial H}}{{\partial n_\varphi  }};\;\dot
\theta  = \frac{1}{\hbar }\frac{{\partial H}}{{\partial n_\theta
}};\;\dot \chi  = \frac{1}{\hbar }\frac{{\partial H}}{{\partial
n_\chi  }}$. The equations of motion for conjugate variables are:
\begin{equation}\label{Conj1}
\dot n_\varphi   =  - \frac{1}{\hbar }\frac{{\partial
H}}{{\partial \varphi }} =  - \frac{{\alpha E_J }}{\hbar }\sin
(\varphi  - 2\theta ) - \frac{{E_J }}{{\hbar \beta }}\left(
{\varphi  - \varphi _X } \right)
\end{equation}
\begin{equation}\label{Conj2}
\dot n_\theta   =  - \frac{1}{\hbar }\frac{{\partial H}}{{\partial
\theta }} =  - \frac{{2\alpha E_J }}{\hbar }\sin (\varphi  -
2\theta ) - \frac{{2E_J }}{\hbar }\sin \theta \cos \chi
\end{equation}
\begin{equation}\label{Conj3}
\dot n_\chi   =  - \frac{1}{\hbar }\frac{{\partial H}}{{\partial
\chi }} =  - \frac{{2E_J }}{\hbar }\cos \theta \sin \chi
\end{equation}
Below we consider Hamiltonian (\ref{Ham2}) as quantum mechanical
with commutator relations imposed on its variables
\begin{equation}\label{Comm1}
\left[ {\varphi ,n_\varphi  } \right] = i;\quad \left[ {\theta
,n_\theta  } \right] = i;\quad \left[ {\chi ,n_\chi  } \right] = i
\end{equation}
\section{Current operator}
From the first principles a current in the loop is equal to the
first derivative of the state energy relative to external flux:
\begin{equation}\label{CurrA}
    I=\frac{\partial E_n}{\partial \Phi_X}
\end{equation}
This expression can be rewritten in terms of exact Hamiltonian of
a system:
\begin{equation}\label{CurrB}
I = \left\langle n \right|\frac{{\partial \hat H}}{{\partial \Phi
_X }}\left| n \right\rangle
\end{equation}
From (\ref{CurrB}) we would make ansatz  that the current operator
is as follows:
\begin{equation}\label{CurrC}
\Hat{I} = \frac{{\partial \hat H}}{{\partial \Phi _X }}
\end{equation}
However (\ref{CurrC}) is not a consequence of (\ref{CurrB}).
Therefore, the ansatz (\ref{CurrC}) must be proved in every case,
since the current operator in the form of Eq. (\ref{CurrC}) has to
be consistent with its definition in terms of variables of
Hamiltonian $H$. The prove for our case is given below.

The current operator across every junction is a sum of a
supercurrent and a current through the capacitor:
\begin{equation}\label{Curr1}
\hat{I}_i  = I_0 \sin \varphi _{_i }  + \frac{\hbar }{{2e}}C\ddot
\varphi _i \quad \left( {i = 1,2} \right)
\end{equation}
\begin{equation}\label{Curr2}
\hat{I}_3  = \alpha I_0 \sin \varphi _{_3 }  + \alpha \frac{\hbar
}{{2e}}C\ddot \varphi _3
\end{equation}
Since the current in a loop is unique the equations (\ref{Curr1})
and (\ref{Curr2}) must give identical result. This is indeed the
case if we express phases $\phi_i (i=1, 2, 3)$ in terms of $\phi$,
$\theta$, $\chi$ and use the equations (\ref{Motion1}),
(\ref{Motion2}), (\ref{Motion3}), (\ref{Conj1}), (\ref{Conj2}),
(\ref{Conj3}). For every $I_i$ in (\ref{Curr1}), (\ref{Curr2}) we
obtain the same expression
\begin{equation}\label{CurrOp1}
\hat{I}  =  - I_0\frac{{\varphi  - \varphi _X }}{\beta }
\end{equation}
which is independent of parameters of a particular junction in the
loop. From the other hand the expression (\ref{CurrOp1}) can be
obtained from our Hamiltonian (\ref{Ham2}) with the aid of
(\ref{CurrC}). Therefore, the  equation (\ref{CurrC}) gives us the
true expression for the current operator. It is important to note
that the proper expression  for the current operator
(\ref{CurrOp1}) cannot be obtained without magnetic energy term in
the original Lagrangian (\ref{Lag1}).

It follows from (\ref{Ham2}) and (\ref{CurrOp1}) that $\left[
{\hat I,\hat H} \right] \ne 0$. Therefore, an eigenstate of $H$
cannot possess a definite current value.

\subsection{Current operator in a two-state basis}
Suppose a system is well described by two low lying states
$|\Psi_\pm\rangle$ with corresponding eigenenergies $E_\pm$:
\begin{equation}\label{Ham3}
    \widehat{H}|\Psi_\pm\rangle=E_\pm|\Psi_\pm\rangle
\end{equation}
Within this subspace Hamiltonian can be expressed in terms of
Pauli matrices $\sigma_X, \sigma_Y, \sigma_Z$:
\begin{equation}\label{Ham4}
    \widehat{H}=\frac{E_++E_-}{2}-\frac{E_+-E_-}{2}\sigma_Z
\end{equation}
with $\sigma_Z|\Psi_\pm\rangle=\mp|\Psi_\pm\rangle$.

Now we calculate the matrix elements of the current operator
(\ref{CurrC}) within this subspace. According to
(\ref{CurrA}),(\ref{CurrB}) and (\ref{CurrC}) diagonal matrix
elements are:
\begin{equation}\label{Currdi}
\left\langle {\Psi _ \pm  } \right|\widehat{I}\left| {\Psi _ \pm }
\right\rangle  = \frac{{\partial E_ \pm  }}{{\partial \Phi _X }}
\end{equation}
In order to find nondiagonal matrix elements of the current
operator we use the expression
\begin{equation}\label{nondiag}
\left\langle n \right|\frac{{\partial \hat H}}{{\partial \lambda
}}\left| {n'} \right\rangle  = \left( {E_{n'}  - E_n }
\right)\left\langle {n\left| {\frac{{\partial n'}}{{\partial
\lambda }}} \right.} \right\rangle
\end{equation}
which is obtained by differentiating of the identity $\left\langle
n \right|\hat H\left| {n'} \right\rangle  = 0$ with respect to
parameter $\lambda$. Hence, the nondiagonal elements of the
current operator are:
\begin{equation}\label{Currnd}
\left\langle {\Psi _ -  } \right|\hat I\left| {\Psi _ +  }
\right\rangle  = \left\langle {\Psi _ +  } \right|\hat I\left|
{\Psi _ -  } \right\rangle  = \left( {E_ +   - E_ -  }
\right)\left\langle {\Psi _ -  \left| {\frac{{\partial \Psi _ +
}}{{\partial \Phi _X }}} \right.} \right\rangle
\end{equation}
Therefore, we can express the current operator in terms of Pauli
matrices:
\begin{equation}\label{CurrOp2}
\widehat I = \frac{\partial }{{\partial \Phi _X }}\left(
{\frac{{E_ + + E_ -  }}{2}} \right)\textbf{I} - \frac{\partial
}{{\partial \Phi _X }}\left( {\frac{{E_ +   - E_ -  }}{2}}
\right)\sigma _Z  + \left( {E_ +   - E_ -  } \right)\left\langle
{\Psi _ -  \left| {\frac{{\partial \Psi _ +  }}{{\partial \Phi _X
}}} \right.} \right\rangle \sigma _X
\end{equation}
where $\textbf{I}$ is the unity matrix.

Below we consider two low lying states of a flux qubit
\begin{equation}\label{En}
E_ \pm   = E_0  \pm \sqrt {\varepsilon ^2  + \Delta ^2 }
\end{equation}
where $E_0$ and the tunneling rate $\Delta$ are independent of the
external flux $\Phi_X$, and the quantity $\varepsilon$ is linear
function of the flux, $\varepsilon=E_J\lambda f_X$, where
$\lambda$ is a numerical factor which depends on qubit parameters
$\alpha$ and $g=E_J/E_C$, $f_X=\Phi_X/\Phi_0-1/2$.

Therefore, for the flux qubit we get in eigenstate basis:
\begin{equation}\label{Ham3}
H = E_0  - \Delta _\varepsilon  \sigma _Z
\end{equation}
\begin{equation}\label{CurrOp3}
\hat I =  - \frac{{\partial \Delta _\varepsilon  }}{{\partial \Phi
_X }}\sigma _Z  + 2\Delta _\varepsilon  \left\langle {\Psi _ -
\left| {\frac{{\partial \Psi _ +  }}{{\partial \Phi _X }}}
\right.} \right\rangle \sigma _X
\end{equation}
where $\Delta _\varepsilon = \sqrt {\varepsilon ^2  + \Delta ^2}
$.

Transformation to the flux basis is obtained via the rotation
around $y$ axes in a two level subspace with the aid of the matrix
$R = \exp \left( {i\xi \sigma _Y /2} \right)$, where $\cos \xi
=\varepsilon/\Delta_\varepsilon$, $\sin \xi
=\Delta/\Delta_\varepsilon$: $R^{ - 1} \sigma _Z R = \tau _Z \cos
\xi  + \tau _X \sin \xi $, $R^{ - 1} \sigma _X R = -\tau _Z \sin
\xi + \tau _X \cos \xi $, where $\tau_X, \tau_Z$ are Pauli
matrices in a flux basis. Hence, we get for Hamiltonian
(\ref{Ham3}) and current operator (\ref{CurrOp3}) in the flux
basis:
\begin{equation}\label{Ham4}
 H =  - \varepsilon \tau _Z  - \Delta \tau _X
\end{equation}
\begin{equation}\label{CurrOp4}
\hat I =  - \left( {\frac{{\partial \Delta _\varepsilon
}}{{\partial \Phi _X }}\frac{\varepsilon }{{\Delta _\varepsilon }}
+ 2\Delta _\varepsilon  \left\langle {\Psi _ -  \left|
{\frac{{\partial \Psi _ +  }}{{\partial \Phi _X }}} \right.}
\right\rangle } \right)\tau _Z  - \left( {\frac{{\partial \Delta
_\varepsilon  }}{{\partial \Phi _X }}\frac{\Delta }{{\Delta
_\varepsilon  }} - 2\varepsilon \left\langle {\Psi _ -  \left|
{\frac{{\partial \Psi _ +  }}{{\partial \Phi _X }}} \right.}
\right\rangle } \right)\tau _X
\end{equation}

Therefore, the current operator is not diagonal neither in the
flux basis nor in the eigenstate basis.

The stationary state wave functions $\Psi_\pm$ can be written as
the superpositions of the wave functions in the flux basis,
$\Psi_L, \Psi_R$ where $L$, $R$ stand for the left, right well,
respectively: $\Psi_\pm=\emph{a}_\pm\Psi_L+\emph{b}_\pm\Psi_R$,
where
\begin{equation}  \label{a,b}
a_\pm=\frac{\Delta}{\sqrt{2\Delta_\varepsilon(\Delta_\varepsilon\mp\varepsilon)}}; b_\pm=\frac{%
\varepsilon\mp\Delta_\varepsilon}{\sqrt{2\Delta_\varepsilon(\Delta_\varepsilon\mp\varepsilon)}};
\end{equation}
The coefficients $a_\pm$, $b_\pm$ are defined in such a way, that
$\tau_Z|\Psi_L\rangle=-|\Psi_L\rangle$,
$\tau_Z|\Psi_R\rangle=+|\Psi_R\rangle$,
$\tau_X|\Psi_L\rangle=+|\Psi_R\rangle$,
$\tau_X|\Psi_R\rangle=+|\Psi_L\rangle$. In terms of the functions
$\Psi_L, \Psi_R$ the cross term $\left\langle {\Psi _ - \left|
{\frac{{\partial \Psi _ +  }}{{\partial \Phi _X }}} \right.}
\right\rangle $ will read
\begin{equation}\label{Crossterm}
\left\langle {\Psi _ -  \left| {\frac{{\partial \Psi _ +
}}{{\partial \Phi _X }}} \right.} \right\rangle  = a_ -
\frac{{\partial a_ +  }}{{\partial \Phi _X }} + b_ -
\frac{{\partial b_ +  }}{{\partial \Phi _X }} + a_ -  a_ +
\left\langle {\Psi _L \left| {\frac{{\partial \Psi _L }}{{\partial
\Phi _X }}} \right.} \right\rangle  + b_ -  b_ +  \left\langle
{\Psi _R \left| {\frac{{\partial \Psi _R }}{{\partial \Phi _X }}}
\right.} \right\rangle
\end{equation}

 The results obtained up till now are exact in
that we did not make any approximation to the Hamiltonian
(\ref{Ham2}). However, in order to calculate $\varepsilon$ and
cross term $\left\langle {\Psi _ - \left| {\frac{{\partial \Psi _
+  }}{{\partial \Phi _X }}} \right.} \right\rangle $ in
(\ref{CurrOp3}),  (\ref{CurrOp4}) we need some approximate
procedure.
\section{Approximation to quantum mechanical Hamiltonian}
In order to calculate the matrix elements of the current operator
we have to find the wave functions of two lowest levels of
Hamiltonian (\ref{Ham2}). First we single out of the potential
(\ref{PotEn}) the fast variable $\varphi$, which describe the
interaction of the qubit with its own $LC$ circuit. The point of
minimum $\varphi_C$ of $U(\chi, \theta, \varphi)$ (\ref{PotEn})
with respect to $\varphi$ is defined from $\partial
U/\partial\phi=0$:
\begin{equation}\label{phiC}
\varphi _C  = \varphi _X  - \alpha \beta \sin \left( {\varphi _C -
2\theta } \right)
\end{equation}
In the vicinity of $\varphi_C$ the potential $U(\chi, \theta,
\varphi)$ can be written as:
\begin{equation}\label{PotEn1}
U\left( {\chi ,\theta ,\varphi } \right) \approx U\left( {\chi
,\theta ,\varphi _C } \right) + \frac{{E_J }}{{2\beta
}}\widehat{\varphi}^2  + \widehat{\varphi}^2\frac{{\alpha E_J
}}{2}\cos \left( {\varphi _C  - 2\theta } \right)
\end{equation}
where $\widehat\varphi$ is a small operator correction to
$\varphi_C$: $\varphi=\varphi_C+\widehat\varphi$.

As is known the potential $U(\chi, \theta, \varphi_C)$ has a
degenerate point at $\Phi_\mathrm{x}=\Phi_0/2$. Assuming $f_X<<1$,
$\beta<<1$ we obtain near this point:
\begin{equation}\label{phiC1}
\varphi _C  = \pi  + 2\pi f_X  - \alpha \beta \sin 2\theta
\end{equation}
From (\ref{CurrOp1}) and (\ref{phiC1}) we find a current operator
in "coordinate" representation:
\begin{equation}\label{CurrCoord}
    \widehat{I}=I_0\alpha \sin 2\theta
\end{equation}
For $U(\chi, \theta, \varphi_C)$ we obtain near degeneracy point
\begin{equation}\label{PotDeg}
U\left( {\chi ,\theta ,\varphi _C } \right) = -2E_J \cos \theta
\cos \chi  + \alpha E_J \cos 2\theta  + \alpha 2\pi f_X E_J \sin
2\theta  - \frac{{\alpha ^2 \beta E_J }}{2}\sin ^2 2\theta
\end{equation}
Below we follow the procedure described in \cite{Green}. At
$f_\mathrm{x}=0$, the potential~(\ref{PotDeg}) has two minima at
$\chi=0$, $\theta=\pm\theta_*$, with $\cos\theta_*=1/2\alpha$
($\theta_*>0$). Tunnelling lifts their degeneracy, leading to
energy levels $E_\pm=E_0 \pm \Delta$. However, at degenerate bias
the current vanishes, forcing one to move slightly away from this
point. In order to find the levels for $|f_\mathrm{x}|\ll1$ we
expand Eq.~(\ref{PotDeg}) near its minima, retaining linear terms
in $f_\mathrm{x}$, $\beta$ and quadratic terms in $\chi,\theta$.
Define $\theta^\mathrm{r/l}_*$ as the minima, shifted due to
$f_\mathrm{x}$ and $\beta$:
\begin{equation}\label{thetaMin}
\theta^\mathrm{r/l}_*=\pm\theta_*+2\pi\!f_\mathrm{x}\frac{1{-}2\alpha^2}{4\alpha^2{-}1}\
\pm \beta \frac{1-2\alpha^2}{2\alpha (4\alpha^2-1)};
\end{equation}
that is, the upper (lower) sign refers to the right (left) well.
The potential energy (\ref{PotDeg}) then reads:
\begin{equation}\label{PotQuadr}
\frac{{U^{r/l}\left( {\chi ,\theta ,\varphi _C } \right)}}{{E_J }}
= U_0^{r/l} + A^{r/l}\widehat{\chi}^2  +
B^{r/l}\widehat{\theta}^{2}
\end{equation}
where the operator correction
$\widehat\theta=\theta-\theta_*^{r/l}$,
\begin{equation}\label{U0}
U_0^{r/l}  =  - \alpha  - \frac{1}{{2\alpha }} \pm 2\pi f_X
\frac{{\sqrt {4\alpha ^2  - 1} }}{{2\alpha }} - \beta
\frac{{4\alpha ^2  - 1}}{{8\alpha ^2 }}
\end{equation}

\begin{equation}\label{A}
A^{r/l} = \frac{1}{{2\alpha }} \mp 2\pi f_X \frac{{2\alpha ^2  -
1}}{{2\alpha \sqrt {4\alpha ^2  - 1} }} + \beta \frac{{2\alpha ^2
- 1}}{{4\alpha ^2 }}
\end{equation}

\begin{equation}\label{B}
B^{r/l} = 2\alpha  - \frac{1}{{2\alpha }} \mp 2\pi f_X
\frac{{2\alpha ^2 + 1}}{{2\alpha \sqrt {4\alpha ^2  - 1} }} +
\beta \left( { - \frac{1}{4} + \frac{5}{2}\alpha ^2  - 2\alpha ^4
} \right)
\end{equation}
Combining (\ref{PotQuadr}) and (\ref{PotDeg}) in (\ref{Ham2}) we
obtain quadratic quantum mechanical Hamiltonian for the flux qubit
in the left and right well near the degeneracy point:
\begin{eqnarray}\label{HamQuant}
H^{r/l} = E_J U_0^{r/l}  + \left[ {\frac{{4\left( {2\alpha  + 1}
\right)}}{\alpha }E_C \hat n_\varphi ^2  + \frac{{E_J }}{{2\beta
}}\hat \varphi ^2 } \right] + \left[ {2E_C \hat n_\chi ^2  + E_J
A^{r/l}\hat \chi ^2 } \right] \\ \nonumber + \left[ {2E_C \hat
n_\theta ^2 + E_J B^{r/l}\hat \theta ^2 } \right] + 8E_C \hat
n_\theta \hat n_\varphi + \frac{{E_J }}{2}C^{r/l}\hat \varphi ^2
\end{eqnarray}
where
\begin{equation}\label{C}
C^{r/l} = \frac{1}{{2\alpha }}\left[ { - 1 \pm 2\pi f_X \left(
{\frac{{2\alpha ^2  - 1}}{{\sqrt {4\alpha ^2  - 1} }} +
\frac{{\sqrt {4\alpha ^2  - 1} }}{\alpha }} \right) + \frac{\beta
}{{2\alpha }}\left( {1 - 2\alpha  - \frac{{4\alpha ^2  -
1}}{\alpha }} \right)} \right]
\end{equation}
The first term in square brackets in (\ref{HamQuant}) is the
Hamiltonian of $LC$ oscillator of the flux qubit, which is
slightly modified by the last term in (\ref{HamQuant}). the next
two terms in square brackets are oscillator Hamiltonians for the
flux qubit variables, $\chi$ and $\theta$, respectively. The
interaction of the $\theta$ degree of freedom with the qubit $LC$
circuit is given by next-to-last term in (\ref{HamQuant}).

Assuming the frequency $(LC)^{-1/2}$ of the qubit $LC$ circuit is
much higher than the junctions frequencies $E_J/\hbar$,
$E_C/\hbar$ we neglect the interaction of the qubit variables,
$\theta$ and $\chi$ with  the qubit $LC$ oscillator. This is
equivalent to the averaging of Hamiltonian (\ref{HamQuant}) over
the ground state of the LC Hamiltonian. Therefore, for the qubit
Hamiltonian we obtain:
\begin{equation}\label{HamQb}
H_{qb}  = \left\langle H^{r/l} \right\rangle  = E_J U_0^{r/l}  +
\frac{1}{2}\varepsilon_0  + \frac{{E_J }}{2}C^{r/l}\left\langle
{\hat \varphi ^2 } \right\rangle  + \left[ {2E_C \hat n_\chi ^2  +
E_J A^{r/l}\hat \chi ^2 } \right] + \left[ {2E_C \hat n_\theta ^2
+ E_J B^{r/l}\hat \theta ^2 } \right]
\end{equation}
where
\[\varepsilon_0  = \left( {\frac{{8E_C E_J }}{\beta }\frac{{(2\alpha +
1)}}{\alpha }} \right)^{1/2} ;\quad \left\langle {\varphi ^2 }
\right\rangle  = \frac{1}{2}\left( {\frac{{8\beta E_C }}{{E_J
}}\frac{{(2\alpha  + 1)}}{\alpha }} \right)^{1/2}\]

Next we confine ourself only to the ground state of (\ref{HamQb})
in either of the wells.
\begin{equation}\label{GrState}
\varepsilon ^{r/l}  = \frac{1}{2}\varepsilon_0  + E_J U_0^{r/l} +
E_J \frac{{C^{r/l} }}{2}\left\langle {\varphi ^2 } \right\rangle +
\frac{{\hbar \omega _\theta ^{r/l} }}{2} + \frac{{\hbar \omega
_\chi ^{r/l} }}{2}
\end{equation}
where
\begin{equation}\label{omchi}
\hbar \omega _\chi ^{r/l}  = E_J \sqrt {\frac{4}{{\alpha g}}}
\left( {1 \mp 2\pi f_X \frac{{2\alpha ^2  - 1}}{{2\sqrt {4\alpha
^2  - 1} }} + \beta \frac{{2\alpha ^2  - 1}}{{4\alpha }}} \right)
\end{equation}
\begin{equation}\label{omtheta}
\hbar \omega _\theta ^{r/l}  = E_J \sqrt {\frac{{4\left( {4\alpha
^2 - 1} \right)}}{{\alpha g}}} \left( {1 \mp 2\pi f_X
\frac{{2\alpha ^2 + 1}}{{2\left( {4\alpha ^2  - 1} \right)^{3/2}
}} + \beta \frac{\alpha }{{4\alpha ^2  - 1}}\left( { - \frac{1}{4}
+ \frac{5}{2}\alpha ^2  - 2\alpha ^4 } \right)} \right)
\end{equation}
The ground state wave functions in left (right) well are as
follows:
\begin{equation}\label{Wave}
\Psi_\mathrm{R/L}=
  \frac{1}{\sqrt{4\pi}}\left(\frac{\hbar \omega_{\chi} ^{r/l}\hbar \omega_{\theta}
  ^{r/l}}{E_C^2}\right)^{1/4}
  \exp\biggl(-\frac{\hbar\omega_\chi^\mathrm{r/l}}{8E_C}\chi^2
  -\frac{\hbar\omega_\theta^\mathrm{r/l}}{8E_C}
  {(\theta{-}\theta_*^\mathrm{r/l})}^2\biggr)\;,
\end{equation}
The tunneling between two wells lifts degeneracy yielding the well
known result for eigenenergies
$E_\pm=(\varepsilon^\mathrm{l}{-}\varepsilon^\mathrm{r})/2\pm
  \sqrt{(\varepsilon^\mathrm{l}{-}\varepsilon^\mathrm{r})^2\!/4+\Delta^2}$,
  which was given above in Eq. (\ref{En}). The Eqs.
  (\ref{GrState}), (\ref{omchi}), (\ref{omtheta}) allows us to
  calculate the numerical factor $\lambda$ in (\ref{En}):
\begin{eqnarray}\label{lambda}
\frac{{\lambda \left( {\alpha ,g} \right)\alpha }}{\pi } = \sqrt
{4\alpha ^2  - 1}  - \sqrt {\frac{\alpha }{g}} \left(
{\frac{{2\alpha ^2  - 1}}{{\sqrt {4\alpha ^2  - 1} }} +
\frac{{2\alpha ^2  + 1}}{{4\alpha ^2  - 1}}} \right) +\\\nonumber
\sqrt{\beta} \sqrt {\frac{\alpha }{g}} \frac{{\sqrt { 8(2\alpha  +
1)} }}{{8\alpha }}\left( {\frac{{2\alpha ^2  - 1}}{{\sqrt {4\alpha
^2 - 1} }} + \frac{{4\alpha ^2  - 1}}{\alpha }} \right)
\end{eqnarray}
The average current in eigenstates $E_\pm$ is calculated from
(\ref{CurrA}):
\begin{equation}\label{CurrEigen}
I_\mathrm{q}=\frac{\partial E_\pm}{\partial\Phi_X}=\pm
I_\mathrm{c}f_\mathrm{x}\frac{\lambda^2(\alpha,
g)}{2\pi}\frac{E_\mathrm{J}}{\sqrt{\varepsilon^2+\Delta^2}},
\end{equation}
If the deviation from the degeneracy point $f_X=0$ is significant
($\varepsilon>>\Delta$) then the current (\ref{CurrEigen}) reduces
to its local value in a particular well $I_q\rightarrow \pm I_C
\lambda(\alpha,g)/2\pi=\partial\varepsilon^{r/l}/\partial\Phi_X$.
\subsection{The matrix elements of the current operator}
As is seen from Eqs. (\ref{CurrOp3}), (\ref{CurrOp4}), it is
necessary to calculate the quantity $\langle\Psi_-|
 \partial\Psi_+/\partial\Phi_X\rangle$. The calculation yields the
 following result:
 \begin{equation}\label{CurrMatrix}
\left\langle {\Psi _ -  \left| {\frac{{\partial \Psi _ +
}}{{\partial \Phi _X }}} \right.} \right\rangle  = \frac{{\partial
\varepsilon }}{{\partial \Phi _X }}\frac{\Delta }{{2\Delta
_\varepsilon ^2 }} + \beta \frac{\pi }{{4\Phi _0 }}\frac{\Delta
}{{\Delta _\varepsilon  }}F\left( \alpha  \right)
\end{equation}
where
\begin{equation}\label{Falpha}
F(\alpha ) = \frac{{\left( {2\alpha ^2  - 1} \right)^2 }}{{4\alpha
\sqrt {4\alpha ^2  - 1} }} + \frac{{\alpha \left( {2\alpha ^2  +
1} \right)}}{{\left( {4\alpha ^2  - 1} \right)^2 }}\left( { -
\frac{1}{4} + \frac{5}{2}\alpha ^2  - 2\alpha ^4 } \right)
\end{equation}
The correction due to inductance (second term in r. h. s. of
(\ref{Falpha})) is usually small, however, it is responsible for
nonzero value of non diagonal matrix elements of the current
operator in the flux basis. In the eigenstate basis Eq.
(\ref{CurrOp3}) transforms to:
\begin{equation}\label{CurrOp5}
\hat I = \frac{{\partial \varepsilon }}{{\partial \Phi _X
}}\frac{1}{{\Delta _\varepsilon  }}\left( { - \varepsilon \sigma
_Z  + \Delta \sigma _X } \right)
\end{equation}
where we neglect the correction due to inductance.

In the flux basis Eq. (\ref{CurrOp4}) transforms to:
\begin{equation}\label{CurrOp5}
\hat I =  - \frac{{\partial \varepsilon }}{{\partial \Phi _X
}}\left( {\frac{{\varepsilon ^2 }}{{\Delta _\varepsilon ^2 }} +
\frac{\Delta }{{\Delta _\varepsilon  }}} \right)\tau _Z  + \beta
\frac{\pi }{{2\Phi _0 }}\frac{{\varepsilon \Delta }}{{\Delta
_\varepsilon  }}F(\alpha)\tau _X
\end{equation}
The Eq. (\ref{CurrOp5}) is the main result of our calculations. It
shows that the current operator in the flux qubit is not diagonal
in the flux basis as it  is implicitly assumed in most of papers
on the subject. The non diagonal term comes from the finite
inductance of the qubit loop. Though for the usual qubit design
this term is relatively small, nevertheless, it might give
noticeable effects for larger values of $\beta$ in the
arrangements when two flux qubit are inductively coupled via a
term $M\widehat{I}_1\widehat{I}_2$ in the Hamiltonian, where $M$
is a mutual inductance between qubit's loops, $\widehat{I}_1$,
$\widehat{I}_2$ are the current operators of the respective
qubits.

In conclusion, we show that the finite loop inductance of a flux
qubit results in additional non diagonal term in the current
operator in the flux basis. The result is important in the
arrangements with magnetic coupling of two or more flux qubits.

\textbf{Acknowledgements}

I thank E. Il'ichev for critical reading of manuscript and
fruitful discussions.

\end{document}